\documentclass{aa}
\usepackage{natbib}
\bibpunct{(}{)}{;}{a}{}{,}

\usepackage[dvips]{graphicx}
\usepackage{times}
\newcommand\ignore[1]{} 
\begin{document}
\title{Reprocessing the Hipparcos data of evolved stars}
\subtitle{III. Revised Hipparcos period-luminosity relationship for galactic 
long-period variable stars.\thanks{Based on observations from the Hipparcos astrometric 
satellite operated by the European Space Agency (ESA 1997)}}

\author{ G. R.~Knapp\inst{1}\and D.~Pourbaix\inst{2,1}\fnmsep\thanks{Research Associate, F.N.R.S., Belgium}\and I.~Platais\inst{3} \and  A.~Jorissen\inst{2}\fnmsep$^{\star\star}$}

\offprints{G. R. Knapp}

\institute{Department of Astrophysical Sciences, Princeton University,
Princeton, NJ 08544, USA; gk@astro.princeton.edu
\and 
Institut d'Astronomie et d'Astrophysique, Universit\'e
Libre de Bruxelles, CP. 226, Boulevard du Triomphe, B-1050
Bruxelles, Belgium; pourbaix,ajorisse@astro.ulb.ac.be
\and
Department of Physics and Astronomy, The Johns Hopkins University, 3400 North Charles Street, Baltimore, MD 21218, USA;imants@pha.jhu.edu
}
\date{Received date; accepted date} 
 
\authorrunning{Knapp et al.}
\titlerunning{P-L relationship for LPVs}

\abstract{
We analyze the $K$ band luminosities of a sample of galactic
long-period variables using
parallaxes measured by the Hipparcos mission.  The parallaxes are in most
cases re-computed from the Hipparcos Intermediate Astrometric Data
using improved astrometric fits and chromaticity corrections.  The $K$ band
magnitudes are taken from the literature and from measurements by COBE,
and are corrected for interstellar and circumstellar extinction.\\
The sample contains stars of several spectral types: M, S and C, and of several
variability classes: Mira, semiregular SRa, and SRb.  We find that the 
distribution of stars in the period-luminosity plane is independent of
circumstellar chemistry, but that the different variability types have
different P-L distributions.  Both the Mira variables and the
SRb variables have reasonably well-defined period-luminosity relationships, 
but with very different slopes.
The SRa variables are distributed between the two classes, suggesting that they
are a mixture of Miras and SRb, rather than a separate class of stars.
{
New period-luminosity relationships are derived based on our revised Hipparcos parallaxes.  The Miras show a similar period-luminosity
relationship to that found for Large Magellanic Cloud Miras by 
\citet{Feast-1989:a}.
}  
\\
The maximum absolute $K$ magnitude of the sample is about $-8.2$ for both Miras
and semi-regular stars, only a little fainter than the expected AGB limit.  We
show that the stars with the longest periods ($P>400$~d) have high
mass loss rates and are almost all Mira variables.
\keywords{stars:carbon -- stars:AGB and post-AGB -- stars:pulsation}
}
\maketitle

\section{Introduction}

This paper investigates the period-luminosity (P-L) relationship, or 
relationships, for long-period variable stars (LPV) in the solar
neighborhood.  P-L relationships for stars in the instability strips of the 
HR diagram have proven to be of fundamental value in using variable stars as 
secondary distance indicators.  Stellar variability is also a powerful
tool for studying stellar structure, chemistry and mass loss 
\citep[e.g. ][]{Wood-1990:a,Willson-2000:a,Alvarez-2001:a}.  The present
paper investigates whether one or more P-L relationships exist for LPVs in the
local region of the Galaxy, using re-derived Hipparcos parallaxes, $K$ magnitude
data from the literature, and $K$-band flux densities and variability data
derived from the COBE DIRBE time-series data \citep{Boggess-1992:a,Hauser-1998:a}.

The P-L relationship for LPVs has been investigated using the Hipparcos
output catalogue \citep{vanLeeuwen-1997} for samples of objects
at the same distance (in the Large Magellanic Cloud \citep{Glass-1981:a,Feast-1989:a,Groenewegen-1996:a,Cioni-2001:a}, and in the 
bulge of the Galaxy \citep{Alard-2001:a}).  The last two papers make
use of the vast data compilation on variable stars available from 
microlensing surveys.  These papers find
P-L relationships for LMC Miras with both oxygen and carbon chemistry, and
relationships
for semi-regular (SR) variables which have the same slope as for Miras but with two
sequences suggested to correspond to the long and short periods which often
coexist in these stars 
{
[\citet{Cioni-2001:a}; see also the discussion by
\citet{Szatmary-1996:a}, \citet{Bedding-1998:b}, and \citet{Zijlstra-2002:a}].
}
In the Galactic Bulge, \citet{Alard-2001:a}
find a P-L relationship for Mira variables but no relation for SR variables.
Given the potential use of these stars as secondary distance indicators (their
luminosities are high, $\ge 3000 L_{\odot}$ and their effective temperatures
low, $\le$ 2500 K, so that the bulk of their flux is emitted at near-infrared
wavelengths, where interstellar extinction is much lower than at optical
wavelengths) the establishment of a well-defined P-L relationship is of 
significant interest.   Further, the P-L relationship establishes the 
mode in which the star is pulsating, and gives information on the effect
of pulsation on mass loss.

Measurements of the P-L relationship in the LMC and the Galactic Bulge 
provide the slope $a$ of the relationship:
\begin{equation}
M_K =  a \log P({\rm days}) +  b \label{Eq:MK}
\end{equation}
and the zero-point $b$ (using the known distances to the LMC and the bulge).
However, models \cite[e.g. ][]{Wood-1990:a,Wood-2000:a} show that the
zero-point absolute luminosity is metallicity dependent.  
Comparisons of the P-L relationship derived for nearby galactic LPVs,
whose absolute distances are known from parallax measurements and for
which there is much more information available on metallicity
and mass loss, are thus worth investigating.  

Establishment of P-L relationships for nearby LPVs is difficult,
however.  Even after Hipparcos \citep{Hipparcos} there are relatively few 
well determined parallaxes for these rare, luminous stars, and their
use requires attention to statistical biases \citep[cf. ][]{Arenou-1999:a}.
Several studies have assumed that the P-L relationship for nearby Miras
has the same slope as found for the LMC Miras \citep{Groenewegen-1996:a,Bedding-1998:a,Whitelock-2000:a,Bergeat-2001:a} and have used this relationship to 
calculate the zero point
for galactic populations. \citet{Barthes-1999:a}, however, using
both direct and statistical parallaxes for a sample of Miras, find a 
different  slope than that seen in the LMC, while \citet{Alvarez-1997:b} identify 
two groups of Miras with different zero points \citep[cf. also ][]{Jura-1992:a}.

Hipparcos provided the first parallaxes for most LPVs for 
which this information is now available.  
However, most of the parallaxes available even from Hipparcos are measured at
low confidence.  Added to this are reduction problems associated
with the very red colors and extreme variability of these objects.
These factors make the evaluation of absolute magnitudes for these objects
quite uncertain.

We have recently undertaken a re-analysis of the Hipparcos astrometric
data for red giant stars with the aim of improving the accuracy of the
parallax information for these objects.  The re-reduction scheme is 
described in detail by \citet{Platais-2003:a} and \citet{Pourbaix-2003:b}.
The present paper uses improved parallax information
from these studies, plus some Hipparcos catalogue parallaxes, to
re-examine the absolute magnitudes of variable AGB stars.  We work
entirely in $K$ magnitudes, using magnitude data from the literature and
from COBE-DIRBE data.  We also incorporate corrections for interstellar
and circumstellar extinction, and use the DIRBE data to examine the $K$ band 
variability of the stars.
The assembly of the data set is described in
the next Section.  Section 3 discusses the distribution of LPVs in the 
P-L plane and investigates their P-L
relationships.  We also examine the maximum AGB luminosity  for both
carbon- and oxygen-rich stars.  Section 4 briefly discusses data on period
variability, from the literature and from the DIRBE data, and discusses the
relationships among the different types of LPVs.  Section 5
discusses Mira variables with long periods and large mass loss rates.  
%The conclusions are given in Section 6.

\section{Data}

\begin{table}[htb]
\setlength{\tabcolsep}{3pt}
\caption[]{\label{Tab:data1}Stars for which a new solution based on the 
Hipparcos observations was derived.  $\varpi$ and $\varepsilon(\varpi)$ are 
the parallax and its error in mas. $\varpi-\varpi_{\rm HIP}$ is the difference
between the revised parallax and the published one, $A_K$ is the total interstellar 
and circumstellar extinction in the $K$ band (see text). $K$ is the 
2.2$\rm\mu m$ apparent $K$ magnitude.  $P$ is the period in days, 'var' is 
the variability type from CGVS, `Chem' is the carbon/oxygen chemistry of the
atmosphere}
\begin{tabular}{llllllllll}
\hline
HIP & GCVS & $\varpi$ & $\varepsilon(\varpi)$ & $\varpi-\varpi_{\rm HIP}$ 
& $A_K$ & $K$ & $P$ & var. & Chem\\ \hline
\end{tabular}
\end{table}

The sample of objects discussed herein was basically defined by the 
available data: we need parallaxes, $K$ magnitudes, periods and  
variability types.  Secondary information
includes the position (which together with the parallax allows the
calculation of interstellar extinction); the 12$\rm \mu m$ flux 
density \citep{IRAS} which, together with the 2$\rm \mu m$ flux density, gives
an estimate for the circumstellar extinction, and the circumstellar
chemistry, since the dust opacity at 12$\rm \mu m$ is strongly dependent on
its composition.

The data are given in Tables \ref{Tab:data1} and \ref{Tab:data2}, where we list: the Hipparcos catalogue 
number; the variable star name; the parallax $\varpi$ in milliarcseconds 
(mas); the parallax error $\varepsilon(\varpi)$ in mas;
{
 the total (interstellar and circumstellar) $K$ band  extinction:
} 
the $K$ magnitude; the period in days; the variability type; and the chemistry 
(O = ``oxygen-rich'', i.e. $n$(O)$>n$(C); C = carbon star and S =
S star).  The sources of these various quantities are discussed below.

\begin{table}[htb]
\caption[]{\label{Tab:data2}Same as Table \ref{Tab:data1} but for SRb
  variables for which the Hipparcos data have not been reprocessed.}
\setlength{\tabcolsep}{0.7mm}
\begin{tabular}{llllllll}\hline
HIP & GCVS & $\varpi$ & $\varepsilon(\varpi)$ & $A_K$ & $K$ & $P$ & Chem. \\
\hline
13064 &  Z Eri  &  4.07  &  0.9  & 0.01 & 0.32 & 80 &  O \\
13384 &  RR Eri  &  2.96  &  1.04  & 0.01 & 0.52 & 97 &  O \\
19116 &  CY Eri  &  2.87  &  1.  & 0.01 & 1.96 & 25 &  O \\
21046 &  RV Cam  &  3.01  &  0.97  & 0.06 & 0.52 & 101 &  O \\
23680 &  W Ori  &  4.66  &  1.44  & 0.07 & -0.25 & 212 &  C \\
24169 &  RX Lep  &  7.3  &  0.71  & 0.01 & -1.40 & 60 &  O \\
28558 &  DP Ori  &  5.34  &  1.73  & 0.01 & 1.05 & 90 &  O \\
28874 &  S Lep  &  3.63  &  0.78  & 0.02 & -0.46 & 89 &  O \\
34922 &  LZ Pup  &  16.46  &  1.27  & 0.02 & -2.49 & 140.6 &  O \\
38406 &  BC CMi  &  7.77  &  0.99  & 0.00 & 0.83 & 35 &  O \\
41664 &  RT Hya  &  3.68  &  1.02  & 0.01 & 0.09 & 290 &  O \\
42502 &  AK Hya  &  5.12  &  0.92  & 0.02 & -0.61 & 75 &  O \\
44050 &  RT Cnc  &  2.94  &  1.11  & 0.01 & 0.20 & 60 &  O \\
55639 &  T Crt  &  5.04  &  1.06  & 0.01 & 1.20 & 70 &  O \\
61022 &  BK Vir  &  5.68  &  1.12  & 0.01 & -0.88 & 150 &  O \\
62611 &  SV Crv  &  2.72  &  0.97  & 0.01 & 1.07 & 70 &  O \\
63642 &  RT Vir  &  7.25  &  1.1  & 0.01 & -1.12 & 155 &  O \\
64569 &  SW Vir  &  7  &  1.2  & 0.01 & -1.87 & 150 &  O \\
66666 &  V744 Cen  &  6  &  0.76  & 0.02 & -0.72 & 90 &  O \\
71644 &  RV Boo  &  2.54  &  0.98  & 0.00 & -0.01 & 137 &  O \\
73213 &  FY Lib  &  2.97  &  1.19  & 0.03 & 0.12 & 120 &  O \\
78574 &  X Her  &  7.26  &  0.7  & 0.01 & 0.67 & 95 &  O \\
81188 &  TX Dra  &  3.52  &  0.56  & 0.01 & 1.38 & 78 &  O \\
82249 &  AH Dra  &  3.39  &  0.7  & 0.01 & 0.57 & 158 &  O \\
87850 &  OP Her  &  3.26  &  0.54  & 0.02 & 0.05 & 120.5 &  O \\
89669 &  IQ Her  &  3.82  &  0.87  & 0.04 & 0.04 & 75 &  O \\
91781 &  V3879 Sgr  &  2.67  &  1.01  & 0.05 & -0.42 & 50 &  O \\
92862 &  R Lyr  &  9.33  &  0.52  & 0.01 & -2.10 & 46 &  O \\
96204 &  V450 Aql  &  3.8  &  0.9  & 0.04 & -0.08 & 64.2 &  O \\
97151 &  V973 Cyg  &  3.6  &  0.57  & 0.02 & 1.49 & 40 &  O \\
100935 &  T Mic  &  3.44  &  1.19  & 0.01 & -1.59 & 347 &  O \\
106642 &  W Cyg  &  5.28  &  0.63  & 0.02 & -1.48 & 131.1 &  O \\
107487 &  AG Cap  &  2.64  &  0.85  & 0.01 & 1.20 & 25 &  O \\
110099 &  UW Peg  &  3.38  &  1.57  & 0.01 & 2.27 & 106 &  O \\
118249 &  S Phe  &  3.93  &  0.87  & 0.03 & 1.56 & 141 &  O \\
\hline
\end{tabular}
\end{table}

\subsection{Parallaxes}

The sample of LPVs analyzed in this work was selected from stars
observed by Hipparcos \citep{Hipparcos}.  This
mission provided accurate astrometric measurements for a sample of some
120\ts000 stars which is approximately complete to a brightness limit
of $V\sim 8$ but also contains many fainter stars.  Included in the
Hipparcos input catalogue were several hundred evolved cool giant stars,
and Hipparcos provided the first measured parallaxes for almost
all of these.  However, because of their relatively large distances, 
variability, red colors and, for some stars, their faint apparent
magnitudes, the parallax errors for these stars are often larger than the
nominal
Hipparcos 1$\sigma$ accuracy of 1 mas.  
We have therefore undertaken
a re-analysis of the Hipparcos Intermediate Astrometric Data (IAD) 
for these objects.  The re-processing scheme contains two components: 
deriving epoch $V-I$ photometry for each IAD using the method described
by \citet{Platais-2003:a}, and then reprocessing the IAD to derive parallaxes
using epoch chromaticity corrections.  This accounts
for the color dependence of the Hipparcos detector, a particularly 
important aspect of the data processing for the very red, and highly
variable, LPVs. 
For the present study, we used three data sets based on these
new reductions: carbon stars -
the requirement of a measured period removes the much less luminous R
type stars from the sample \citep[cf. ][]{Knapp-2001:a}, leaving the cooler,
more luminous, variable N-type carbon stars; Variability-Induced
Movers \citep[VIMs;][]{Pourbaix-2003:b};
and oxygen-rich Miras. The parallaxes and their errors are given in 
Table \ref{Tab:data1}. Note that in many cases the parallax is not significant.

The samples of carbon stars, VIMs and oxygen-rich Miras defined above are 
complete in the sense that they comprise all such objects observed by 
Hipparcos except for the small number without measured periods or
$K$ magnitudes.  The Hipparcos carbon star and VIM samples contribute a fair
number of SR 
variables: to round out the data sets, we extracted from the Hipparcos
catalogue data for the remaining oxygen-rich SRb observed by Hipparcos.  The
data for this sample, with the same information as in Table \ref{Tab:data1}, 
are given in Table 2.

\subsection{Periods and Variability Types}

The monumental Combined General Catalogue of Variable Stars \citep[CGCVS, ][]{ 
Kholopov-1998:a} contains data for many types of
variable stars, including amplitudes at visual wavelengths, periods,
and classification.  LPVs are classified into several subtypes: 
\underbar{Miras}, with long, well-established, repeatable periods
(typically 300~d or longer) and large amplitudes ($\Delta V
> 2.5$ magnitudes): \underbar{semiregular variables} with shorter,
less repeatable periods and smaller amplitudes, which are subclassified 
into types SRa and SRb: and \underbar{irregular} variables, type Lb,
which show no well-defined period \citep[in fact, periods have been
measured for a few of these, and their variability behavior
is indistinguishable from that of SRb; ][]{Kerschbaum-2001:a}. 
Many semiregular variables
show two or more periods, and some switch between two or more periods.  We
discuss these variability types further in Section 5.

\begin{figure}
\resizebox{\hsize}{!}{\includegraphics{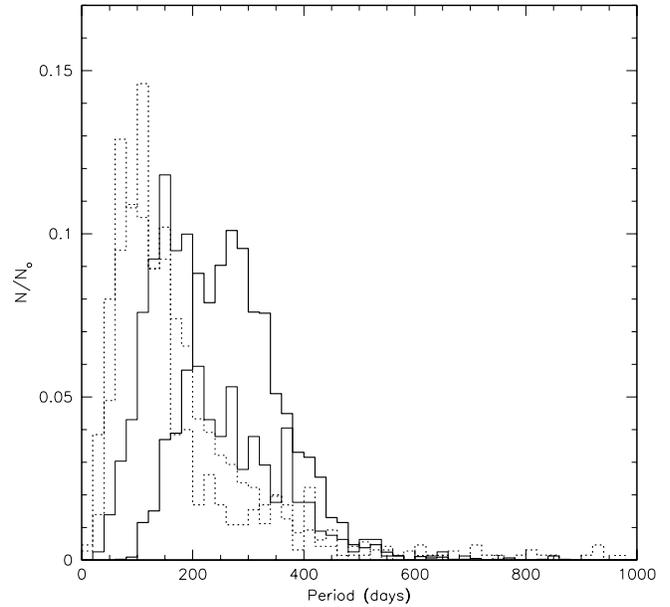}}
\caption{\label{fig1}Frequency of periods from the CGCVS.  Heavy solid line:
Mira.  Light solid line: SRa.  Heavy dotted line: SRb. Light dotted line:
SR.}
\end{figure}
 
Figure 1 shows the period distribution for LPVs from the CGCVS. Many
of the semiregular variables are classified only as SR.  The period 
distributions for all types cover a similar range, and there are few
stars of any type with periods longer than 400 days (we discuss this
further in Section 5).  The median period of the Miras is more than
twice that of the SRb.  The stars classified only as SR have a
period distribution closest to the SRb.  The SRa distribution is 
intermediate between the Miras and SRb.  In this paper, we will
use three classes: Mira, SRa and SRb, combining the last with the SR
and the very small number of stars classified as Lb for which a 
period has been found.  

The periods and variability types in Table \ref{Tab:data1} were taken from the CGCVS and the Hipparcos variability annex.  Periods measured by Hipparcos agree well with the CGCVS periods in almost all cases.

\subsection{$K$ magnitudes}

We analyzed the P-L relation for LPVs using $K$ magnitudes, which are
available for almost all bright LPVs from the literature.  The 
variation of Miras between maximum and minimum light can exceed 10 magnitudes
at visual wavelengths, since the visual luminosity is a highly
non-linear function of photospheric temperature because of molecule
formation \citep[for a recent discussion see ][]{Reid-2002:a}; while
the variability amplitude is much smaller (about 0.5 magnitudes) at $K$
\citep[e.g. ][]{Whitelock-2000:b}.
The $K$ luminosity is also less affected by chemical composition \citep{Wood-2000:a}
and is near the peak of the stellar spectral energy distribution.

$K$ magnitudes were taken from the literature, including the IRC \citep{IRC}, 
the compilation by \citet{CaInOb}, \citet{Kerschbaum-1994:a}
and \citet{Whitelock-2000:b}. Except for the last reference,
which gives $K$ magnitudes averaged over the light curve, 
these are single-epoch
observations.

\begin{figure}
\resizebox{\hsize}{!}{\includegraphics{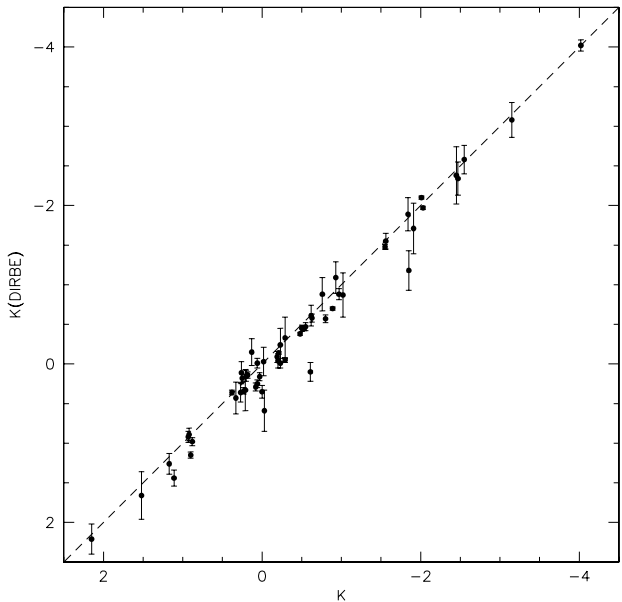}}
\caption{$K$ magnitude from DIRBE 2.2 $\rm \mu m$ flux densities vs.
$K$ magnitude from the literature.  The dashed line shows equality.  The
bars on the DIRBE magnitudes show the observed range of 
magnitude variation over the time span of the COBE observations.}
\label{fig2}
\end{figure}

To study the variance of the $K$ magnitudes caused by variability, we 
extracted 2.2$\rm \mu m$ light curves from the data from the Diffuse
Infrared Background Explorer \citep[DIRBE - ][]{Hauser-1998:a} on the Cosmic
Background Explorer COBE \citep{Boggess-1992:a}.  DIRBE provides well-sampled
partial light curves (sometimes with several observations per day) for
bright point sources at wavelengths from 1.25$\rm \mu m$ to 240$\rm \mu m$
over a time interval of almost 300 days (11 December 1989 to 21 September
1990). The gaps in the data and the short duration of the COBE mission
relative to the typical period of an LPV limit the usefulness of COBE
for measuring periods \citep[see e.g. ][]{Smith-2002:a}.  Because of the 
large DIRBE beam ($40'$) the data are useful only for the brightest
point sources, those brighter than about 0-1 magnitudes.

The DIRBE 2.2$\rm \mu m$ light curves were extracted for the brightest 
objects in Table \ref{Tab:data1} (approximately $0$~mag and brighter at $K$) in the
present sample using software written by N. Odegard and D. Leisawitz 
and available at 
\leftline{\bf http://cobe.gsfc.nasa.gov/cio/browser.html}.
The light curves of some of these stars will be discussed in Section 4.
Figure 2 shows the DIRBE $K$ magnitudes versus those from the literature.
The DIRBE 2.2$\rm \mu m$ magnitudes were calculated from the observed
flux densities assuming a value for the flux density of a $0$~mag star
of 654 Jy [using the calibrations of a $0.03$~mag star by \citet{Bessell-1988:a}].

The DIRBE time series data were used to calculate an average flux density for
the entire data set plus the mean deviation from that average.  In most cases
most of the deviation is found to be due to stellar variability rather than 
noise.  The ranges in magnitudes for the COBE data are also shown in Figure 2.

\begin{figure}
\resizebox{\hsize}{!}{\includegraphics{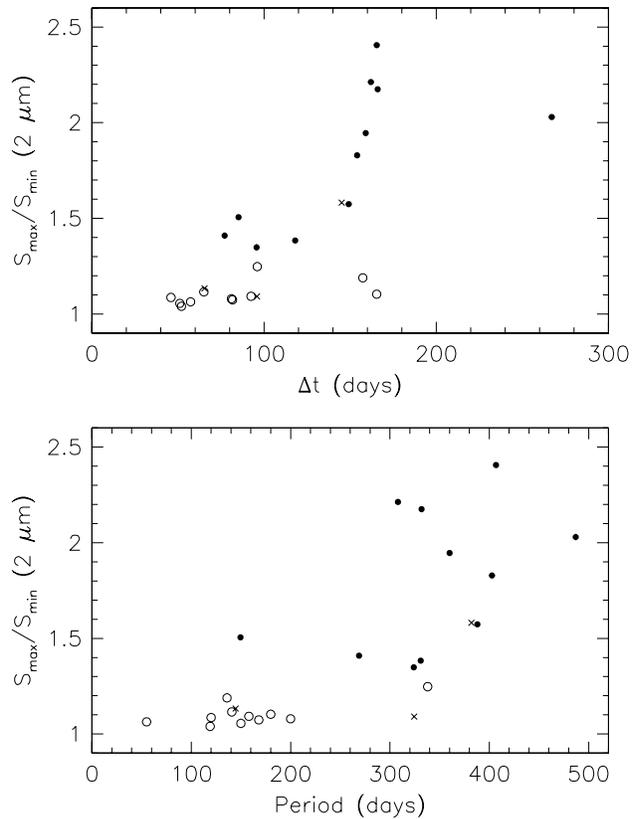}}
\caption{Ratio of maximum and minimum 2.2 $\rm \mu m$ flux densities 
versus (a) time between maximum and minimum 2.2 $\rm \mu m$
light, as measured by DIRBE (upper panel)
and (b) period from the CGCVS (lower panel).
Filled circles: Mira variables.  Open circles: SRb and SR variables.  Crosses:
SRa variables.}
\label{fig3}
\end{figure}

In general the agreement between the data from the literature and from DIRBE
is excellent. The mean internal deviation from the COBE data is 0.2-0.4
magnitudes for Mira variables and about 0.02 - 0.1 mag for SR and SRb
variables. In almost all cases the difference between the DIRBE
and the literature magnitude is within these deviations, which therefore 
dominate the measurement noise.  The data for a few stars show larger 
discrepancies, and may be due to longer-term variations in these stars.

Figure 3 shows the ratio of the maximum to minimum 2$\rm \mu m$ flux density
for stars for which a sufficient portion of the light curve is observed by 
DIRBE to observe at least one maximum and one minimum (only a minority
of the stars in Table \ref{Tab:data1}).  Figure 3a shows this
ratio versus the time difference between the minimum and maximum, or
vice-versa - not necessarily half the period, since the light curves
for LPVs are often asymmetric.  Figure 3b shows the ratio versus the
period from the CGCVS.  Figure 3 shows that Miras are far more 
readily distinguished from SRb by amplitude than by  period \citep[cf. also
 ][]{Cioni-2001:a}, although stars with $(t_{\rm max}-t_{\rm min})$ shorter than 
about 70 days are always SRb.  The SRa variables overlap both the Miras and 
SRb and are therefore likely to be a mixture of the
two types (cf. Figure 1).

In the next section, then, we discuss only two samples: Miras
and SR/SRb.  We do not include the SRa in the analysis.  We assume that
the error in the COBE median $K$ magnitude is $0.1$ mag for SR and $0.3$ mag
for Miras,in both cases dominated by variability; these are the median values
found from the COBE light curves.

\subsection{Extinction}

The extinction towards the stars has two components, interstellar and
circumstellar. We used the model of \citet{Hakkila-1997:a}
to calculate $A_V$, the total interstellar extinction in $V$, from the position of each star and its distance, estimated from $\varpi^{-1}$.
The $K$ band extinction is then found from $A_{K_i}=0.114\; A_V$ \citep{Cardelli-1989:a}.

Many of the stars in Table \ref{Tab:data1} are losing mass and surrounded by 
dusty circumstellar envelopes.  \citet{Ivezic-1997:a} show that the
resulting spectral energy distribution is a function of the dust 
composition  and the optical depth.  The circumstellar extinctions
for the stars discussed in this paper are not large by the standards
of mass-losing AGB stars (they are all visible objects) and the optical
depth can be estimated from the ratio of the 12$\rm \mu m$ flux density
(emitted almost entirely by the dusty envelope) to the 2$\rm \mu m$
flux density (mostly emitted by the photosphere).  The relationship
between the 12$\rm \mu m/2 \mu m$ flux ratio and the extinction at
$K$ was estimated using a spherically symmetric dusty envelope model
\citep{Knapp-1993:a} using silicate grains for
oxygen and S stars \citep[cf. ][]{Jorissen-1998:b} and
carbonaceous grains for carbon stars.  The model assumed condensation
temperatures of 800~K and 1000~K for silicate and carbonaceous grains
respectively \citep[cf. ][]{Ivezic-1997:a} to define the radius
of the inner edge of the model envelope.  The model's central star is a 
blackbody with $T_{\star}=2500$~K, $L_{\star}=5000$~L$_{\odot}$, 
for which the intrinsic ratio is $S_{12}/S_{2.2}= 0.18$.  For low optical depths
\begin{equation}
A_{K_c} \approx0.07\left({S_{12}\over S_{2.2}} - 0.18\right)
\end{equation}
for silicate envelopes, and
\begin{equation}
A_{K_c}\approx1.1\left({S_{12}\over S_{2.2}}-0.18\right)
\end{equation}
for carbonaceous dust.  The large difference between the two is because of the
strong contribution of the silicate 9.7$\rm \mu m$ emission feature in the 
IRAS 12$\rm \mu m$ band. A very small number of stars have no IRAS 
observations: we assume a circumstellar extinction of 0 mag for these stars.

{
The total (interstellar + circumstellar) extinction at $K$ band is listed in Tables \ref{Tab:data1} and \ref{Tab:data2}.
}
The uncertainties in both the interstellar and circumstellar extinction are
large, but the contribution of these errors to the errors in the 
computed absolute $K$ magnitude are dwarfed by those in the parallax and 
by stellar variability, and thus will be neglected in the remainder
of this paper.

\section{The Period-Luminosity relationship}

\begin{figure}
\resizebox{\hsize}{!}{\includegraphics{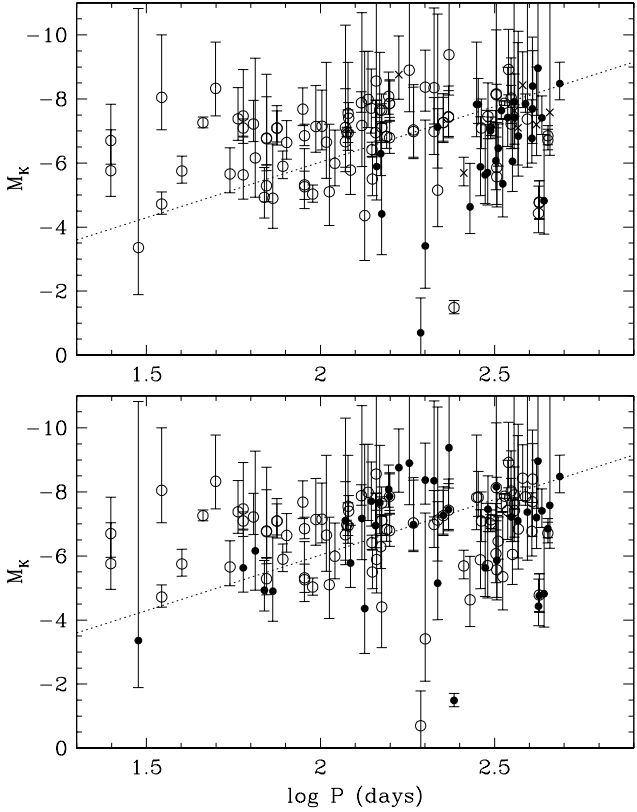}}
\caption{\label{Fig:MKlogP}
Distribution of LPVs in the $M_K - \log P$ plane.  (a) Upper panel: stars are sorted by variable type.
Filled circles: Miras.  Open circles: SR/SRb.  Crosses: SRa.
(b) Lower panel: stars are sorted by chemical type.  Filled symbols:
carbon stars.  Open symbols: oxygen stars.  The dashed line is the P-L relationship of Eq.~(\ref{Eq:MK2}).}
\end{figure}

The stars listed in Tables \ref{Tab:data1} and \ref{Tab:data2} often have 
poorly measured parallaxes.  Figure \ref{Fig:MKlogP} shows the distribution
in the period-$M_K$ plane of stars with reliable $\varpi$, defined by:

\begin{itemize}
\item probability that $\varpi>0$ at the 95\% confidence
level.  The mean Hipparcos parallax error is $\sim$ 1 mas with 
some small scatter around this value depending on the location of the star
on the sky and on its brightness. LPVs are very luminous, so many 
observed by Hipparcos may be at distances  $>$ 1 kpc.  The 
probability that the parallax is detected is determined by including/not 
including the parallax in the astrometric solution, and defining
the probability that the parallax is detected using the F-test
\citep[cf. ][]{Pourbaix-2003:b}.
\item $\varpi/\varepsilon(\varpi) > 2$

\item $\varpi > 0.$

\item $\varepsilon(\varpi)<2$ mas. As discussed by \citet{Platais-2003:a} the errors on the Hipparcos parallaxes for some relatively bright red giant stars are anomalously large, and measurements with large errors often produce inaccurate parallaxes.

\end{itemize}

The distance is calculated from $D = 1000/\varpi$ and the absolute
magnitude $M_K$ from this distance and the observed $K$ magnitude,
corrected for circumstellar and interstellar extinction. The errors
on $M_K$ are assumed to be due to the errors (or rather range
of photometric variation) in $K$ and the parallax error.  In \ref{Fig:MKlogP},
the stars
are coded by variability type (Mira, SR/SRb and SRa) and by circumstellar
chemistry (combining the oxygen and S stars).   Also shown in Figure 
\ref{Fig:MKlogP}
is the period-luminosity relationship for LMC Miras by \citet{Feast-1989:a}
with the zero point as determined for galactic Miras
by \citet{Whitelock-2000:b}:
\begin{equation}
M_K  =  -3.47 \log P({\rm days}) + 0.85\label{Eq:MK2}
\end{equation}

Figure \ref{Fig:MKlogP}  shows several features:

\begin{enumerate}

\item
To first order, the location of LPVs on the P-L diagram is a scatter plot,
as shown also by \citet{Bedding-1998:a} for oxygen Miras
and by \citet{Bergeat-2001:a} for carbon stars.

\item
The Miras and SR are partly-separated on this diagram \citep[cf. also][]{Bedding-1998:a}.  The Miras have longer periods (no
star with a period shorter than 120 days is a Mira), but
both types of variables have similar luminosity ranges.

\item
The separation is not complete.  Several SR are located in the Mira
region, and vice versa.

\item
The pulsation characteristics of carbon and oxygen-rich Miras are
indistinguishable.

\item The SR P-L sequence, if there is one, has a different slope from that of the Miras.

\item
The partial overlapping of the Mira and SRb regions implies that some 
Miras may switch to SRb and vice-versa. As we discuss in the next section, there is ample evidence for mode-switching between Mira-like
and SR-like pulsation modes, and the incomplete segregation in Figure \ref{Fig:MKlogP} may be due to this. 

\item
The (relatively small) number of SRa stars are spread approximately 
equally between the Mira and SRb regions of the diagram.

\item Both the Mira and SR variables have a maximum luminosity at $K$ of $M_K=-8.2\pm0.2$.  The maximum luminosity also holds for both oxygen and carbon stars, and is in agreement with the value found for LMC stars.

\end{enumerate}

\subsection{The Period-Luminosity relationship}

We now use the data presented in Figure \ref{Fig:MKlogP} to calculate the mean 
P-L relation.  We discuss the Mira and SRb samples
separately and do not include the SRa.

Many of the objects do not have significant parallaxes, even after reprocessing
the Hipparcos data, and absolute magnitudes derived from the subset
of objects in Tables \ref{Tab:data1} and \ref{Tab:data2} and depicted in Figure \ref{Fig:MKlogP}
which are selected by 
parallax (and/or by $\varpi/\varepsilon(\varpi$)) will be biased \citep{Lutz-1973:a}.  The correct estimation of mean absolute 
magnitudes is then analogous to weighting the ``detections'' of 
parallax by the upper limits on the parallax on the assumption that
the stars with detected and non-detected parallaxes are drawn from the 
same population.  While the Hipparcos catalogue is magnitude
limited and therefore absolute magnitudes are also subject to Malmquist bias,
the parallax for an LPV may become undetectable with respect to stars of
similar brightness before the star does - a 
star with $M_K = -8$ at $D = 1$ kpc has $V\sim 8$ - and errors on the
parallaxes are therefore the major source of uncertainty.

{\bf
Although the Astrometry Based Luminosity \citep[ABL, ][]{Arenou-1999:a} is an unbiased estimator of the luminosity, it cannot be used directly to calculate absolute magnitudes which can then be used to derive a period-luminosity relation: the symmetry of the error bars on ABL does not propagate to the absolute magnitude.  Instead of fitting the absolute magnitudes, we work in parallax space, where the errors are Gaussian.  Therefore, instead of fitting Eq.(\ref{Eq:MK}), we fit
\begin{equation}
\varpi=P^{a\over 5}10^{b\over 5}10^{10-K+A_K\over 5}
\end{equation}
where $\varpi$, the parallax in mas, can be used regardless of its sign.  The parameters are fitted by chi-square minimization using non-linear optimization technique, and their uncertainties are obtained by mean of a standard Monte Carlo simulation.
}

\begin{figure}
\resizebox{\hsize}{!}{\includegraphics{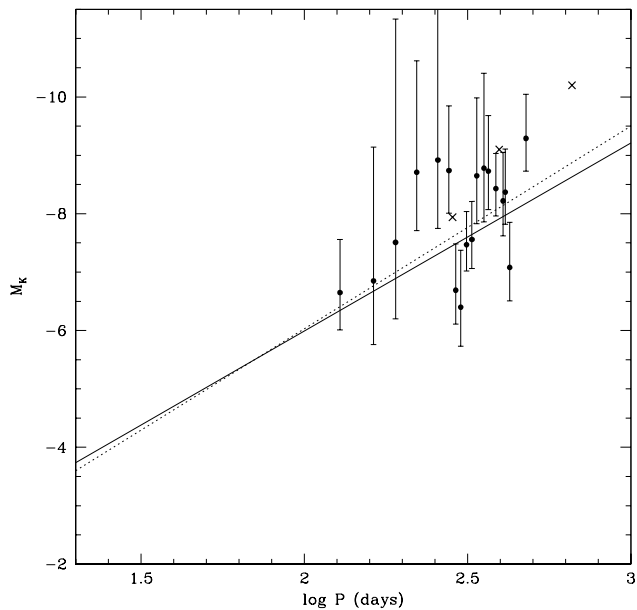}}
\caption{Binned P-L data for Mira variables.  The LMC relationship
of \citet{Feast-1989:a} is shown by the dotted line, and the 
least-squares fit to the data by the solid line.  Crosses: oxygen-rich
Mira variables with distances from OH phase-lag measurements (see text).}
\label{plmira}
\end{figure}

%\begin{figure}
%\resizebox{\hsize}{!}{\includegraphics{plsr.eps}}
%\caption{Binned distribution of $M_K$ vs. $\log P$ for
%SRb/SR variables.  The dotted line shows the relation of \citet{Feast-1989:a} for LMC Miras, while the solid line is a least-squares fit
%to the data.  Filled circles: SR/SRb variables from Table \ref{Tab:data1}.
%Crosses: SR/SRb variables from Table 2.}
%\label{plsr}
%\end{figure}

Figure \ref{plmira} shows the resulting mean $M_K-\log P$ diagram for Miras.  The data in that figure are strongly correlated: the
linear correlation coefficient is 0.84, giving a $\gg 99$\% chance
of correlation.
The least-squares fit to the P-L relationship for Miras gives
\begin{equation}
M_K = -3.22(\pm0.38)\; \log P ({\rm days}) +  0.45 (\pm2.40) \label{Eq:MK3}
\end{equation}
{\bf
 which is very consistent with that found by \citet{Feast-1989:a} and is shown in Fig.~\ref{plmira}.
}

Figure \ref{plmira} also shows P-L data for three nearby OH/IR stars
with reliable phase-lag distances \citep{vanLangevelde-1990:a}:
WX Psc (IRC+10011), RR Aql and R Aql.  The last object is nearby enough to
be measured by Hipparcos: the phase lag distance, 290 pc, is in 
reasonably good agreement with the Hipparcos parallax ($\varpi^{-1}
 = 235\pm52$ pc).  
{
These data also show reasonable correspondence with the derived P-L slope.
}

The data for the SR variables also show a correlation,
though less strongly: the correlation coefficient is 0.65, giving a $>$90\%
chance of correlation.  The least-squares fit is:
\begin{equation}
M_K  =  -1.34(\pm0.06)\; \log P ({\rm days}) - 4.5(\pm0.35)\label{Eq:MK5}
\end{equation}

%The above results disagree both with those of \citet{Feast-1989:a}
%for the LMC (except for stars with periods longer than 400 days), and

{\bf
The above results disagree with
with those of \citet{Barthes-1999:a} who divide their data into different
samples, finding different relations for each sample, with slopes shallower
than the LMC Mira slope.  
}

\section{Variability and Mass Loss}

Miras with periods as long as nearly 2000 days have been found.  These
are all OH/IR stars with large mass loss rates.  As discussed below, mass loss becomes very high for long-period variable stars,
completely obscuring the star at visible and near-infrared wavelengths.
Thus a large part of the period space occupied by LPVs is inaccessible
to studies of the P-L relationship because parallax values 
for these stars are measured at present only at optical wavelengths.
However, there are indications that the P-L relationships so far found
do not hold for very long period Miras.  First, if it did, the predicted 
bolometric luminosity for a star with a $\rm 2000$~d period would be
$\rm 7 \times 10^4\; L_{\odot}$, well in excess of the maximum AGB luminosity
of $\rm 5 \times 10^4\; L_{\odot}$ \citep{Paczynski-1970:a}.
Recently, \citet{He-2001:a} have examined the P-L relationship
for OH/IR stars with phase-lag or kinematic distances; there is a large amount
of scatter in the P-L diagram, and the relationship may be different for 
objects in different IRAS spectral classes.  Furthermore, the bolometric 
luminosities of many of these highly obscured, very long period stars
are relatively modest, often $\rm < 10^4\; L_{\odot}$.

\begin{figure}
\resizebox{\hsize}{!}{\includegraphics{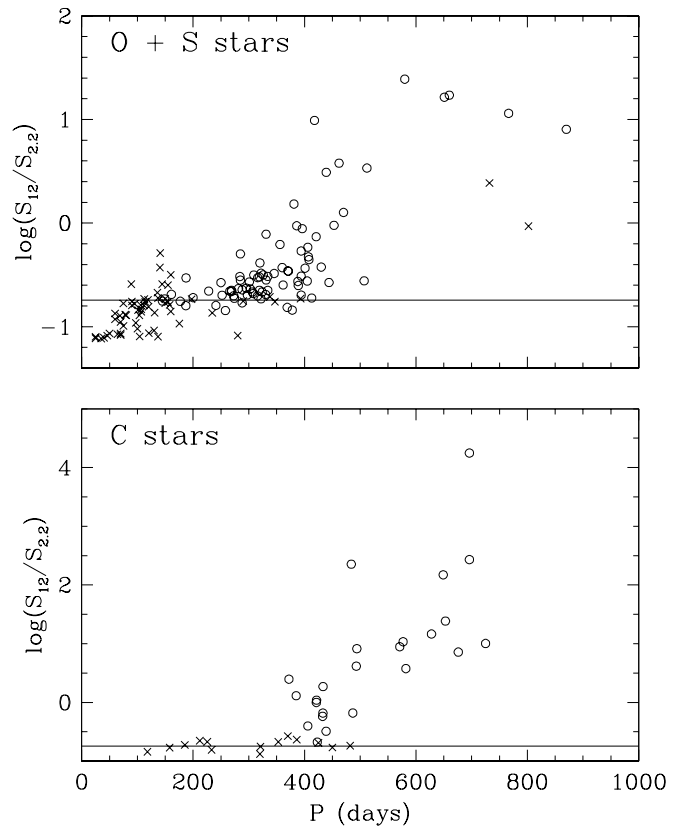}}
\caption{Infrared color ($\rm S_{12}/S_{2.2}$) versus period for oxygen
stars (upper panel) and carbon stars (lower panel).  Open symbols:
Mira variables. Crosses: SR/SRb variables.  The horizontal line shows the 
$\rm S_{12}/S_{2.2}$ for a 2500 K black body.}
\label{mloss}
\end{figure}

The CGCVS contains very few stars with periods 
longer than 400 days.  However, such stars are known, for example
the OH/IR stars. \citet{LeBertre-1993:a,LeBertre-1992:a} measured near infrared light
curves for several tens of oxygen-rich and carbon stars which were first 
discovered in infrared sky surveys.  These objects are heavily obscured,
and almost all of them have periods longer than 400 days.  Almost all of them
vary with large amplitudes and are therefore Mira variables.
In Figure \ref{mloss}, the IR colors of these stars 
are compared with those from Tables \ref{Tab:data1} and 2, showing the
ratio of the 12$\rm \mu m/2.2\rm \mu m$ flux densities versus the period
(as noted in Section 2, this quantity is a good indicator of mass loss).
The objects plotted are those from Tables \ref{Tab:data1} and \ref{Tab:data2} for which we have the 
appropriate measurements, plus the stars 
for which \citet{LeBertre-1993:a,LeBertre-1992:a}
measured periods (with the exception of the OH/IR stars,
which are often at large distances and therefore have 
an unknown amount of interstellar
extinction), and, to round out the samples, the carbon star
CIT 6 [period from \citet{Alksnis-1995:a}]
and the oxygen supergiants VY CMa (Mira) and VX Sgr (SRb). 

Figure \ref{mloss} shows that copious mass loss  turns on more-or-less
abruptly at periods of about 300 - 350 days for both oxygen and carbon stars.
Further, the mass loss rate is
much higher for Miras at a given period than it is for SR variables
(the flattening in $S_{12}/S_{2.2}$ for oxygen/S stars at long
periods is due to the saturation of the 12$\rm \mu m$ flux as the 
9.7 $\rm \mu m$ silicate feature goes into absorption with increasing
optical depth). 
Stars with high mass loss rates are
essentially invisible at optical wavelengths.  Thus the P-L 
relation is incomplete: stars with periods longer than 
about 400 days are absent.  This part of period space will only become
accessible by astrometry at the few microarcseconds level or better at infrared wavelengths.

\section{Variability modes in LPVs}

\begin{figure}
\resizebox{\hsize}{!}{\includegraphics{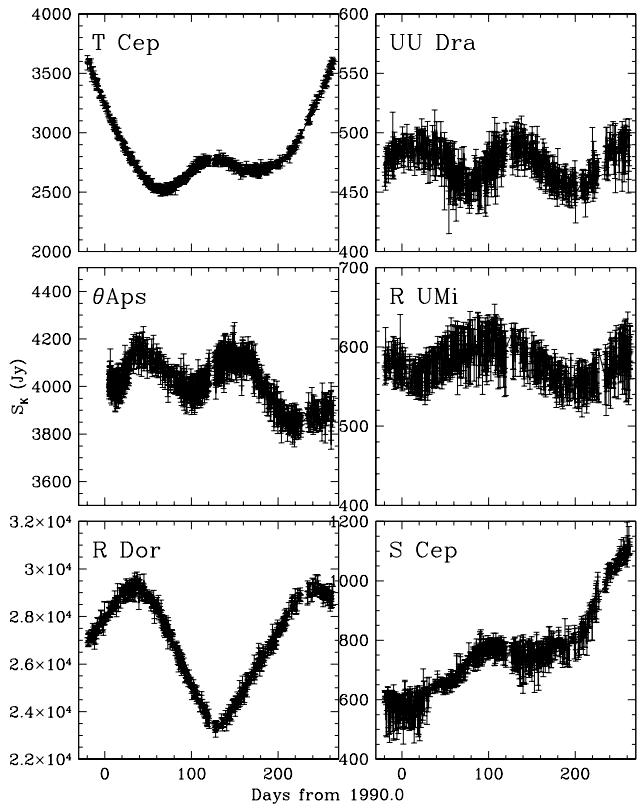}}
\caption{Example COBE $K$ band light curves for LPVs.  T Cep and
S Cep are Miras, R UMi is SRa, and the other 
stars are SR.  A small number of data
points with large variance have been removed. 
{\bf
Both T cep and S Cep show the bump on the rising part of the light curve often seen for Miras \citep{Lockwood-1971:a}
}
}
\label{lcurv}
\end{figure}

Figure \ref{lcurv} illustrates some additional problems with studying the P-L
relation for LPVs.  It shows DIRBE $2.2\rm\mu m$
light curves for six stars for which  almost complete time coverage is 
available: two Miras (S Cep and T Cep), one SRa (R UMi) and three SRb
($\theta$ Aps, UU Dra and R Dor).  The SR variables all have periods short
enough that a complete phase information is available, and the SRa, R UMi,
is indistinguishable in its variability characteristics from the other SR
stars.  Both Mira variables have periods that are longer than the COBE time 
coverage of 300 days: T Cep's period is $\rm 388$~d, S Cep's $488$~d.
The partial COBE light curves are consistent with this period.

Table \ref{Tab:DIRBE} lists period data from DIRBE, where available, and from the literature, for stars with multiple periods, all of which are in the 
sample analyzed earlier in this paper.  \citet{Kerschbaum-2001:a}
note that fewer than 50\% of the periods for SRb stars determined from their 
monitoring observations agree with those from the CGCVS: the data in 
Table \ref{Tab:DIRBE} support this conclusion.

\begin{table}[htb]
\caption{\label{Tab:DIRBE}Periods}
\setlength{\tabcolsep}{0.7mm}
\begin{tabular}{lllll}\hline
HIP & GCVS & var & P & Ref \\ 
\hline
  21479 & R Dor  &  SRb & 209     &  DIRBE\\
        &       &       & 175,332 &  \citet{Bedding-1998:a}\\
        &       &       & 338     &  CGCVS\\
  28166 & BQ Ori&   SRa & 127,240 &  \citet{Kiss-1999:a}\\
        &       &       & 110     &  CGCVS\\
  36288 & Y Lyn &   SRc & 133,205 &  \citet{Szatmary-1996:a}\\
        &       &       & 110,1400&  \citet{Percy-2001:a}\\
        &       &       & 110     &  CGCVS\\
  53085 & V Hya &   SRa & 530,6000&  \citet{Knapp-1999:a}\\
        &       &       & 529     &  CGCVS\\
  63950 & FS Com&   SRb & 55, 600 &  \citet{Percy-2001:a}\\
        &       &       & 58      &  CGCVS\\
  65835 & R Hya &    M  &(314)    &  DIRBE\\
        &       &       & 388     &  CGCVS\\
  67419 & W Hya &   SRa & 274     &  DIRBE\\
        &       &       & 382     &  CGCVS\\
  68815 & $\theta$ Aps& SRb & 121     &  DIRBE\\
        &       &       & 119     &  CGCVS\\
  70969 & Y Cen &   SRb &(338)    &  DIRBE\\
        &       &       & 180     &  CGCVS\\
  80802 & R UMi &   SRa & 181     &  DIRBE\\
        &       &       & 170,325 &  \citet{Kiss-1999:a}\\
        &       &       & 324     &  CGCVS\\
  81747 & AX Sco&   SR  & (39),(64)& \citet{Kerschbaum-2001:a}\\
        &       &       & 124,128  & \citet{Kerschbaum-2001:a}\\
        &       &       & 138     &  CGCVS\\
  81835 & S Dra &   SRb & 182     &  DIRBE\\
        &       &       & 136     &  CGCVS\\
  95173 & T Sge &   SR  & (112)   &  \citet{Kerschbaum-2001:a}\\
        &       &       & 178,316 &  \citet{Kerschbaum-2001:a}\\
        &       &       & 166     &  CGCVS\\
 100605 & UU Dra&   SRb & 117     &  DIRBE\\
        &       &       & 120     &  CGCVS\\
 104451 & T Cep &    M  & 112,$>$290& DIRBE\\
        &       &       & 388     &  CGCVS\\
\hline
\end{tabular}
\end{table}

Several analyses of very long time series data for individual stars have 
recently appeared in the literature.  \citet{Bedding-1998:b}
find two periods, 175~d and 332~d, for R Dor, with switching 
between the two periods on a timescale of about 1000 days. The DIRBE
light curve has an intermediate period of 209~d, intermediate
between the long and short periods.  \citet{Bedding-1998:b}
note that the longer period, plus the star's well-determined absolute 
magnitude, place it on the LMC Mira P-L relation (note that it also 
lies on the Mira P-L relation found above) and suggest that the
star switches pulsation modes between SR and Mira variability, i.e.
between third and first overtone radial pulsation, with Miras
considered to be pulsating in the first overtone \citep{Wood-1990:a,Willson-2000:a} - although see the discussion by \citet{Barthes-1998:a}.

%The mode switching seen in such stars as R Dor may also be present in S Cep
%and T Cep (cf. Figure \ref{lcurv}), illustrating 
As pointed out by 
\citet{Lebzelter-2001:a} understanding LPVs requires both short-
and long-term monitoring.  Other studies of long time-series for LPVs
include that of \citet{Howarth-2001:a} who find two periods
for the Mira variable T Cas, 445~d and 222~d, whose ratio suggests
that the longer period is the fundamental \cite[cf. ][]{Barthes-1998:a};
switching between these two modes takes place on a time scale of about 
3000~d. \citet{Szatmary-1996:a} and \citet{Bedding-1998:b}
find that V Boo appears to have changed from a 
Mira to an SR variable.  \citet{Cadmus-1991:a} also find
evidence of mode switching for several stars.
{\bf
\citet{Sterken-1999:a} show that the period of $\chi$ Cyg is increasing.
}
One star which does
not appear to undergo such behavior is V Hya, for which \citet{Knapp-1999:a}
find repeatable and well established periods
of $530$~d and $6000$~d.  Finally, \citet{Marengo-2001:a}
examine the structure of circumstellar shells due
to mass loss, in which the time-history of that mass loss may be seen, and
suggest that essentially all LPVs undergo mode switching between the
Mira and semi-regular phases.

\section{Discussion and Conclusions}

In this paper, we have examined the distribution of LPVs
in nearby regions of the Galaxy in the $P - M_K$ plane 
using distances derived from Hipparcos parallaxes.  We find P-L relationships 
for both Mira and SRb variables.  The slope for the
SRb is much shallower than that for the Miras.  The relationships
found herein do not agree with some previous relationships found in the literature.
{
 The P-L relationship for Miras agrees well with that found in the LMC.

We discuss information from the present paper and from the literature
on period changes in LPVs and on mass loss and its dependence on the
variability period.  The evidence suggests that LPVs,
at least those with periods $\rm < 400$~d, change variability modes
back and forth between Mira and SR variation, with models suggesting that
different radial pulsation modes become dominant.
}  
However, the P-L relationships found here do not extend to periods longer than
about 700 days.

If visible LPVs, i.e. those with low mass loss rates, switch between the
Mira and SR modes, the shallowness of the P-L relationship for SRb
relative to that for Miras is easy to understand: the period of a 
given star may decrease during the Mira $\rightarrow$ SRb phase by
a factor of 2-4 (depending on the pulsation modes) but the luminosity
stays about the same.

We also show that mass loss is strongly dependent on period and
variability mode.  Miras at a given period are losing much more mass than 
are SR, and the mass loss rate rises steadily with period for both
Miras and SRb.  Stars with periods longer than about 400 - 500 days
are losing so much mass that they become highly obscured at visible
wavelengths and have no parallax information: thus a large part of 
period space is not available for studies of the P-L relations.  These
very long period variables are almost all Miras, suggesting that
the mode switching characteristic of shorter-period red variables disappears
as the period lengthens.

Comparing the P-L relationships for LPVs from the literature and this
paper, it appears that the results are strongly dependent on the
sample selection.  The use of P-L relationships to determine distances
to LPVs is therefore not likely to be useful in at least the near future.  
However, the maximum $K$ absolute magnitude in the nearby (few 100 pc)
region of the Galaxy is $-8.2\pm0.2$, in good agreement with observations
of the LMC.  This corresponds to a bolometric luminosity of about
5000 - 6000 $\rm L_{\odot}$, about the maximum possible luminosity for 
AGB stars. This result, in itself, shows the unlikeliness of a P-L relationship
which extends to periods longer than 400 - 500 days: stars with shorter
periods are already close to the AGB luminosity limit.

Further progress on studying the P-L characteristics of LPVs would be
greatly aided by astrometric measurements at the few micro-arcsecond
level in near-infrared bands, which would make possible the analysis 
of the entire period range.

\begin{acknowledgements}
{
We thank the referee, F. van Leeuwen, and M.~Feast for many valuable comments.
}
We are very grateful to NASA for generous support of this work via grant
NAG5-11094 as well as to ESA via PRODEX C15152/01/NL/SFe(IC). This research 
made use of the SIMBAD database, operated at
CDS, Strasbourg, France; of the NASA Astrophysics Data System, operated at
the Harvard-Smithsonian Center for Astrophysics; of the astro-ph electronic
preprint server operated by Cornell University; and of the data analysis
and display program SM, written by Robert Lupton and Patricia Monger.
We used archival data from the Hipparcos mission, operated by the
European Space Agency, and from the COBE mission, operated by NASA via
the Goddard Space Flight Center.

\end{acknowledgements}

\bibliographystyle{aa} 
\bibliography{articles,books}

\end{document}